\documentclass[twocolumn,showpacs,preprintnumbers,amsmath,amssymb]{revtex4}

\usepackage{graphicx}
\usepackage{dcolumn}
\usepackage{bm}

\begin{document}

\title{Peculiarities of superconducting properties of thin superconductor-normal metal bilayer with large ratio of
resistivities}

\author{D. Yu. Vodolazov}

\email{vodolazov@ipmras.ru}

\affiliation{Institute for Physics of Microstructures, Russian
Academy of Sciences, 603950, Nizhny Novgorod, GSP-105, Russia}

\affiliation{Physics Department, Moscow State University of
Education, Moscow, Russia}

\author{E. E. Pestov}
\affiliation{Institute for Physics of Microstructures, Russian
Academy of Sciences, 603950, Nizhny Novgorod, GSP-105, Russia}
\affiliation{Lobachevsky State University of Nizhny Novgorod, 23
Gagarin Avenue, 603950 Nizhny Novgorod, Russia}

\author{S. N. Vdovichev}
\affiliation{Institute for Physics of Microstructures, Russian
Academy of Sciences, 603950, Nizhny Novgorod, GSP-105, Russia}
\affiliation{Lobachevsky State University of Nizhny Novgorod, 23
Gagarin Avenue, 603950 Nizhny Novgorod, Russia}

\author{M. Yu. Levichev}
\affiliation{Institute for Physics of Microstructures, Russian
Academy of Sciences, 603950, Nizhny Novgorod, GSP-105, Russia} v

\author{S. S. Ustavshikov}
\affiliation{Institute for Physics of Microstructures, Russian
Academy of Sciences, 603950, Nizhny Novgorod, GSP-105, Russia}
\affiliation{Lobachevsky State University of Nizhny Novgorod, 23
Gagarin Avenue, 603950 Nizhny Novgorod, Russia}

\author{A. Yu. Aladyshkin}
\affiliation{Institute for Physics of Microstructures, Russian
Academy of Sciences, 603950, Nizhny Novgorod, GSP-105, Russia}
\affiliation{Lobachevsky State University of Nizhny Novgorod, 23
Gagarin Avenue, 603950 Nizhny Novgorod, Russia}

\author{A. V. Putilov}
\affiliation{Institute for Physics of Microstructures, Russian
Academy of Sciences, 603950, Nizhny Novgorod, GSP-105, Russia}

\author{P.A. Yunin}
\affiliation{Institute for Physics of Microstructures, Russian
Academy of Sciences, 603950, Nizhny Novgorod, GSP-105, Russia}

\author{A. I. El'kina}
\affiliation{Institute for Physics of Microstructures, Russian
Academy of Sciences, 603950, Nizhny Novgorod, GSP-105, Russia}

\author{N. N. Bukharov}
\affiliation{Institute for Physics of Microstructures, Russian
Academy of Sciences, 603950, Nizhny Novgorod, GSP-105, Russia}

\author{A. M. Klushin}
\affiliation{Institute for Physics of Microstructures, Russian
Academy of Sciences, 603950, Nizhny Novgorod, GSP-105, Russia}

\date{\today}

\begin{abstract}

We demonstrate, both theoretically and experimentally, that thin
dirty superconductor-normal metal bilayer with resistivity of
normal metal $\rho_N$ much smaller than normal-state resistivity
of superconductor $\rho_S$ has unique superconducting properties.
First of all the normal layer provides the dominant contribution
to the diamagnetic response of whole bilayer structure in wide
temperature interval below the critical temperature due to
proximity induced superconductivity. Secondly, the presence of the
normal layer may increase the critical current $I_c$ in several
times (the effect is not connected with enhanced vortex pinning),
provides strong temperature dependence of both $I_c$ and effective
magnetic field penetration depth even at temperatures much below
the critical one and leads to the diode effect in parallel
magnetic field. Besides of general interest we believe that the
found results may be useful in construction of different kinds of
superconducting detectors.

\end{abstract}

\maketitle

\section{Introduction}

If superconductor (S) is attached to the normal metal (N) and the
SN interface is transparent for electron motion then
superconducting electrons penetrate the normal metal on the
characteristic length scale $\xi_N(T)$. It leads to
superconducting properties of the normal metal, namely, it can
carry non-dissipative current and screen applied magnetic field.
The screening effect was observed experimentally in many works on
different SN systems \cite{Oda,Bergmann,Mota} at temperatures much
below the critical temperature of the superconductor $T_{c0}$ and
the theory of this effect was developed both in clean and dirty
limits using Eilenberger or Usadel equations
\cite{Zaikin,Pambianchi,Belzig,Fauchere,Galaktionov}.

Here we demonstrate that thin low resistive normal (N) layer
placed on the superconducting (S) layer with large normal-state
resistivity $\rho_S$ can considerably increase its superconducting
properties, namely considerably enhances the diamagnetic response
and critical current $I_c$. We argue that the effect comes from
proximity induced superconductivity and locally smaller London
penetration depth $\lambda$ in N-layer - see inset in Fig. 1(a).
In this system the critical current enhancement is not connected
with enhanced vortex pinning in SN bilayer but it is related with
large superconducting current density $j_s\sim 1/\lambda^2$ in
N-layer. The strong temperature dependence of $I_c$ even at $T\ll
T_c$, suppression of $I_c$ in rather weak perpendicular magnetic
field, the diode effect in parallel magnetic field and enhanced
diamagnetism validates in favor of this interpretation of the
found experimental results.

We have to mention that the critical current density $j_{cN}$ in
N-layer of dirty SNS trilayer with thickness of N-layer $d_N \ll
\xi_N(T)=(\hbar D_N/k_BT)^{1/2}$, thickness of S-layer $d_S\gg
\xi_c=(\hbar D_S/k_BT_{c0})^{1/2}\sim \xi(0)$ ($D_{S(N)}$ is a
diffusion coefficient of corresponding layers, $T_{c0}$ is the
critical temperature of superconductor, $\xi(0)$ is the zero
temperature superconducting coherence length) and ratio of
resistivities $\rho_S/\rho_N \gg 1$ first was analytically
calculated in Refs. \cite{Aslamazov,Lempitskii}. Authors predicted
that $j_{cN}$ may exceed depairing current density $j_{dep}$ of
the superconductor and $j_{cN}(T)\sim 1/\sqrt{T}$ but they did not
calculate the critical current of whole structure. Our
calculations show that N-layer could lead to enhancement of
critical current $I_c$ of whole bilayer and nontrivial $I_c(T)$
only at low temperatures $T \ll T_{c0}$ when thickness of S or N
layers exceed several $\xi_c$ (for bilayer with realistic
$\rho_S/\rho_N \lesssim 200$) while in wide temperature interval
nontrivial $I_c(T)$ exists when $d_S, d_N \lesssim 2\xi_c$.
Because in Refs. \cite{Aslamazov,Lempitskii} the symmetric SNS
system was studied the diode effect in parallel magnetic field was
absent.

The structure of the paper is following. In Sec. II we present our
theoretical results. In Sec. III we show results of the experiment
and in Sec. IV we compare our experimental and theoretical
results, discuss their relation with other experiments and
possible application of such bilayers.

\begin{figure}[hbtp]
\includegraphics[width=0.49\textwidth]{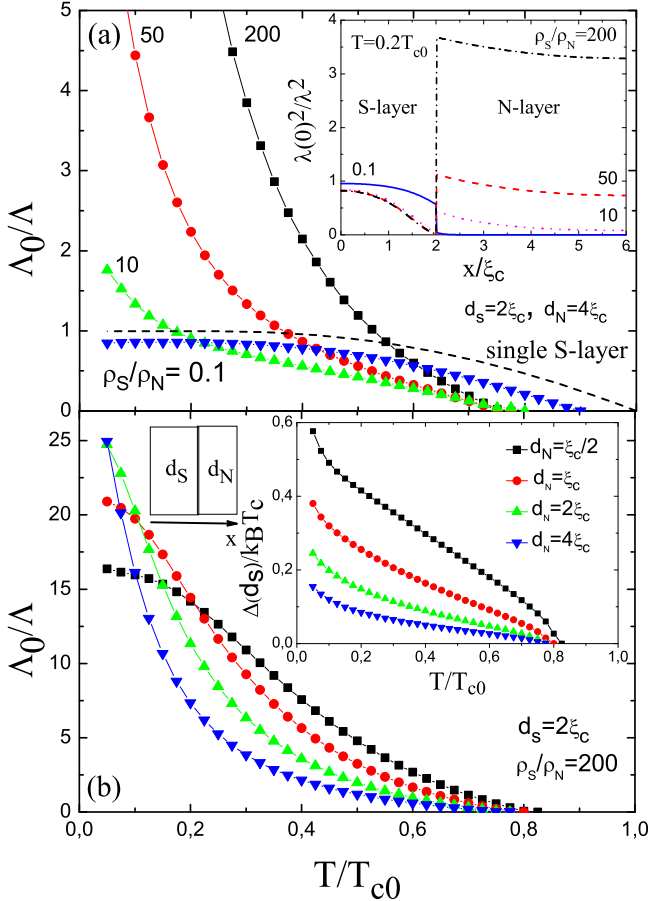}
\caption{Temperature dependence of inverse effective magnetic
field penetration depth $\Lambda^{-1}$ of SN bilayer at different
$\rho_S/\rho_N=200, 50, 10, 0.1$  (a) and different thicknesses of
N-layer $d_N/\xi_c=1/2, 1, 2, 4$ (b). In inset to Fig. 1(a) we
show dependence of local $1/\lambda^2$ across the thickness of
bilayer. In S-layer $1/\lambda^2$ is strongly suppressed near SN
interface due to inverse proximity effect. In inset to Fig.1(b) we
present dependence $\Delta(d_S)(T)$ ($\Delta$ near SN interface)
to demonstrate its correlation with $\Lambda^{-1}(T)$.}
\end{figure}

\begin{figure}[hbtp]
\includegraphics[width=0.55\textwidth]{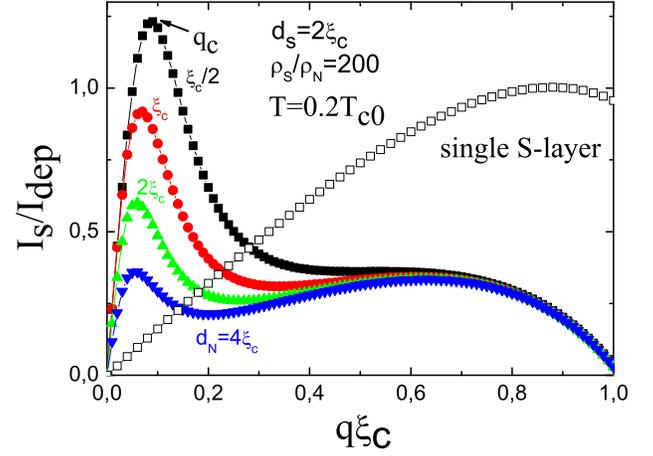}
\caption{Dependence of the superconducting current $I_s$ (it is
normalized to depairing current of single S-layer) flowing along
the bilayer on $q$ which is proportional to supervelocity $v_s
\sim q$. Results are present for different thicknesses of N-layer
$d_N/\xi_c= 1/2, 1, 2,4$. Critical current corresponds to maximal
possible value of $I_s$ and for bilayers it is reached at
$q\xi_c\sim 0.06-0.09$ for chosen parameters. In single S-layer
maximum in dependence $I_s(q)$ (empty squares) is located at
$q\xi_c\simeq 0.88$.}
\end{figure}

\section{Theoretical results}

To calculate superconducting properties of SN bilayer we mainly
use Usadel equation for anomalous $F$ and normal $G$ Green
functions (equations and details of numerical calculations are
present in Appendix A). In Fig. 1(a,b) we show calculated inverse
effective magnetic field penetration depth $\Lambda^{-1}=\int
dx/\lambda(x)^2$ (in a case of single superconducting layer with
$\lambda(x)=const$, $\Lambda=\lambda^2/d_S$ - is the Pearl
penetration depth \cite{Pearl}). $\Lambda$ is normalized in units
of $\Lambda_0=\lambda(0)^2/d_S$, where $\lambda(0)$ is the London
penetration depth of single S-layer at $T=0$. $\Lambda^{-1}$
describes the screening ability of bilayer and for relatively
large ratio $\rho_S/\rho_N$ and temperature not far below from
$T_c$ it could considerably exceeds $\Lambda^{-1}$ of single
S-layer (see Fig. 1(a)). This effect originates from relation
$1/\lambda^2\sim 1/\rho$ valid in the dirty limit and proximity
induced superconductivity in N-layer (see inset in Fig. 1(a)).
Besides one can see non-BCS (Bardeen-Cooper-Shreefer) like
temperature dependence of $\Lambda^{-1}$ which is consequence of
the temperature dependence of superconducting order parameter
$\Delta(d_S)$ near the SN interface (see inset in Fig. 1(b)) which
controls the strength of induced superconductivity in N-layer.
Value of $\Delta(d_S)$ depends not only on the temperature and
ratio of resistivities but also on the thickness of N-layer. We
consider the situation when $d_N<\xi_N(T_{c0})$ and even small
variation of $d_N$ or temperature leading to small change of ratio
$d_N/\xi_N(T)$ strongly influences $\Delta(d_S)$ and
$\Lambda^{-1}$ due to large parameter $\rho_S/\rho_N\gg 1$ in the
boundary condition for anomalous Green function $F$ at SN
interface: $dF/dx|_{d_S-0}=(\rho_S/\rho_N)dF/dx|_{d_S+0}$ and
boundary condition $dF/dx|_{d_S+d_N}=0$ at outer edge.
\begin{figure}[hbtp]
\includegraphics[width=0.48\textwidth]{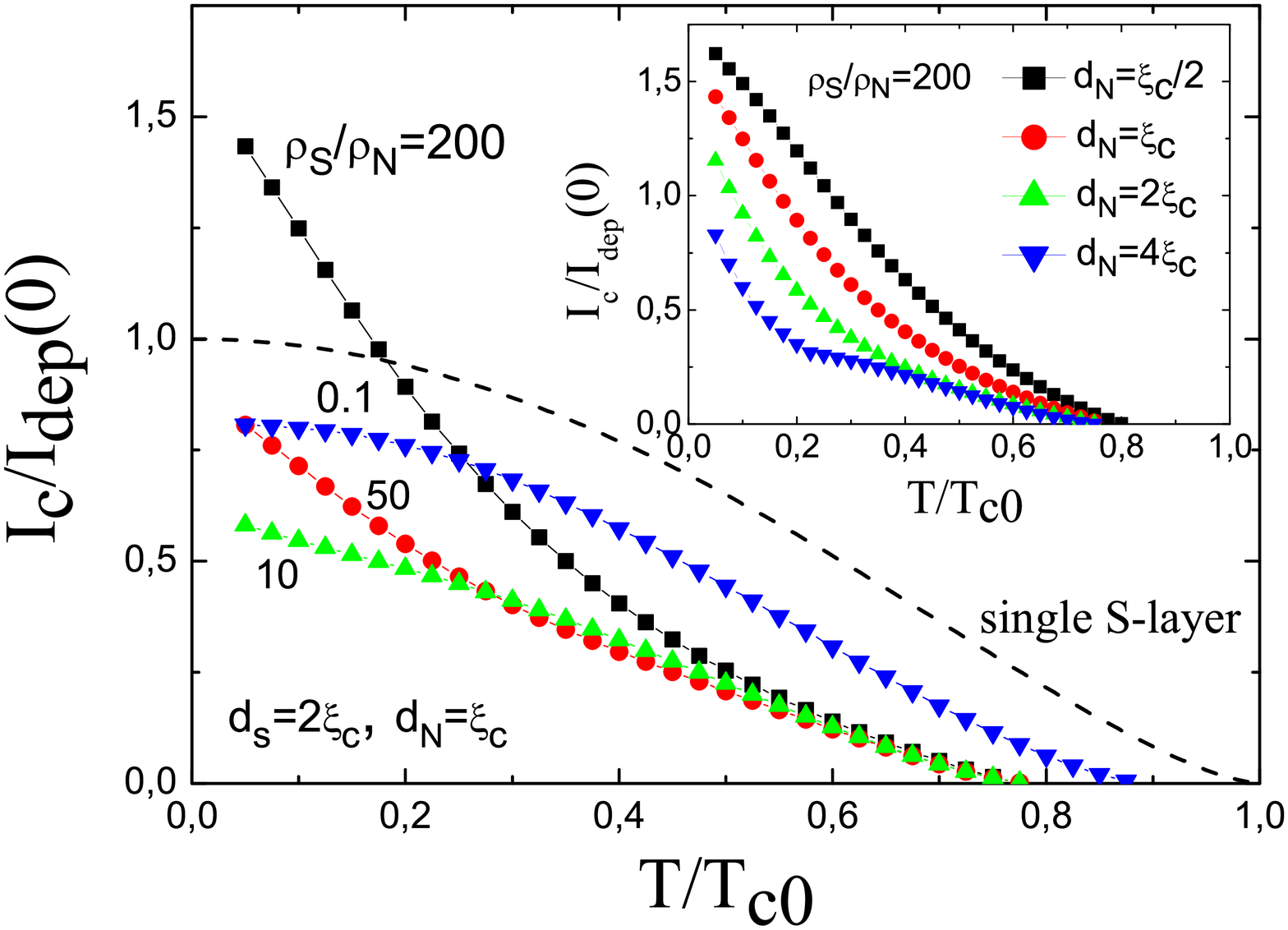}
\caption{Temperature dependence of the critical current of SN
bilayer at different $\rho_S/\rho_N=0.1, 10, 50, 200$ (dashed
curve  corresponds to temperature dependence of depairing current
of single S-layer). In the inset we present temperature dependence
of the critical current of SN bilayer at different thicknesses of
N-layer $d_N/\xi_c= 1/2, 1, 2, 4$ and $d_S=2\xi_c$. For bilayer
with $d_N/\xi_c=4$ dependence $I_c(T)$ has a kink at $T\simeq 0.22
T_{c0}$ because of dominant contribution of N-layer in $I_c$ at $T
\lesssim 0.22 T_{c0}$.}
\end{figure}

In Fig. 2 we present calculated dependence of superconducting
current $I_s$ flowing along the bilayer as a function of value
$q=\nabla \phi-2\pi A/\Phi_0$ ($\phi$ is a phase of
superconducting order parameter, A is a vector potential, $\Phi_0$
is the magnetic flux quantum) which is proportional to the
superconducting velocity $v_s\sim q$ and those value does not vary
across the bilayer (when parallel/perpendicular magnetic field
$H_{||, \bot}=0$). In calculations we assume no vortices and
uniform current distribution across the superconductor/bilayer.
Note that this dependence may have two maxima in contrast with
single superconducting film (see for example \cite{Romijn}). At
$q<q_c$ (see Fig. 2) the major part of superconducting current
flows via the normal layer (due to locally larger $1/\lambda^2$)
while at $q>q_c$ proximity-induced superconductivity in N-layer is
suppressed ($\lambda^{-2}$ rapidly decreases with increasing
$q>q_c$) and near the second maxima the major part of
superconducting current flows via S-layer. For relatively small
thickness of the N-layer the maximal (critical) superconducting
current can exceed the depairing current of single superconducting
film.
\begin{figure}[hbtp]
\includegraphics[width=0.48\textwidth]{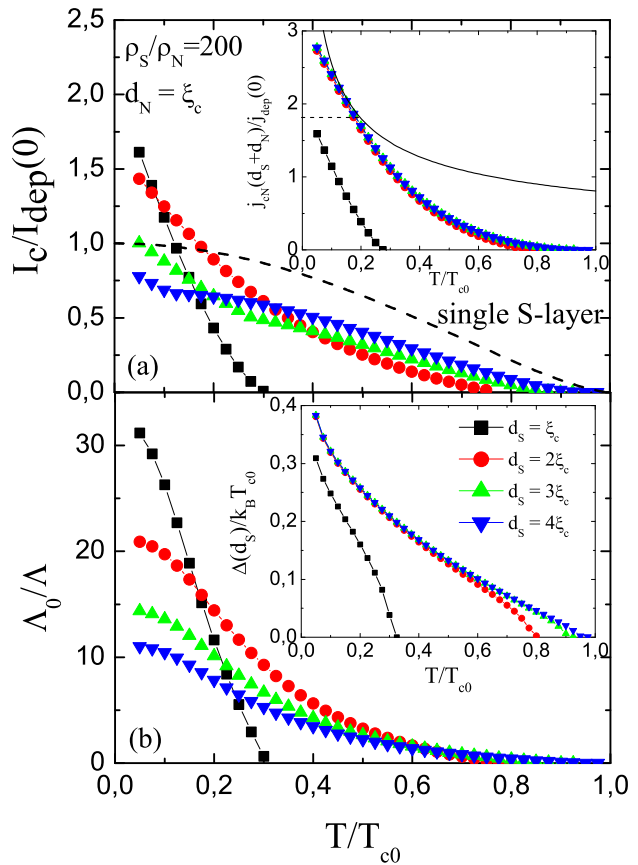}
\caption{Temperature dependence of $I_c$ (a) and $\Lambda^{-1}$
(b) of SN bilayer with different thicknesses of S-layer. With
increasing $d_S$ nontrivial temperature dependence of $I_c$ shifts
to lower temperatures where superconducting current mainly flows
in N-layer. In inset to figure (a) we show temperature dependence
of critical (maximal) superconducting current density in N-layer
at outer boundary ($x=d_S+d_N$) and analytical expression for
$j_{cN}(T)$ found in Ref. \cite{Aslamazov} (solid curve is Eq.
(28) and dashed line is Eq. (30) from \cite{Aslamazov}). In inset
to figure (b) we show temperature dependence of $\Delta$ at SN
interface ($\Delta(d_S)$).}
\end{figure}

Temperature dependence of critical current of bilayer also is
non-BSC like (see Fig. 3) and resembles temperature dependence of
$\Lambda^{-1}$. Origin of this dependence comes from larger
magnitude of proximity induced superconductivity in the N-layer
when temperature decreases and corresponding increase of
$\Lambda^{-1}$.

So far we present results for fixed thickness of S-layer
$d_S=2\xi_c$. We find that the nontrivial (non BCS like)
temperature dependence $I_c(T)$ shifts to lower temperatures with
increasing $d_S$ - see Fig. 4(a) and in wide temperature interval
$(0 \div T_c)$ it exists only for relatively thin bilayers with
$d_S \lesssim 2\xi_c$ (for realistic ratio $\rho_S/\rho_N \lesssim
200$). The reason is in increasing contribution of S-layer to the
superconducting current when $d_S$ increases while contribution of
N-layer stays practically the same. Similar effect occurs with
increasing $d_N$ (at fixed $d_S$, $T$ and not very low
temperatures) because of weaker induced superconductivity in
N-layer. In inset to Fig. 4(a) we present calculated temperature
dependent critical (maximal) superconducting current density in
N-layer and its comparison with analytical results from Ref.
\cite{Aslamazov} (solid and dashed curves - see Eq. (28) and (30),
correspondingly, in \cite{Aslamazov}). Note qualitative similarity
and quantitative difference.

In contrast, $\Lambda^{-1}(T)$ is non BCS like in wide temperature
interval even for relatively large $d_S$ - see Fig. 4(b). The
reason for this is following. Critical current of bilayer is
proportional to $q_c\Lambda^{-1}$, where $q_c$ corresponds to
maximum in dependence $I_s(q)$. In bilayer with $\rho_S/\rho_N \gg
1$ $q_c$ is much smaller than in superconducting film (it roughly
scales as $\sim \rho^{1/2}$). It results only to a little larger
$I_c$ in bilayer despite much larger $\Lambda^{-1}$ in comparison
with superconducting film. As $\Lambda_0/\Lambda$ decreases (for
example with increasing $d_S$) the critical current is mainly
determined by S-layer (except at very low $T$) while main
contribution to $\Lambda^{-1}$ still comes from N-layer when $q
\ll q_c$.

Due to difference in critical supervelocities of S and N-layers
value of critical current in 'positive' ($I^+$) and 'negative'
($I^-$) directions are different in parallel magnetic field
($I^{\pm} \bot H_{||}$ - see inset in Fig. 5). Indeed, parallel
magnetic field either increases $q=\nabla \phi-2\pi A/\Phi_0$ in
N-layer or decreases it depending on direction of the current
(which is determined by direction of $\nabla \phi$). In the first
case $I_c$ rapidly decreases because $q$ reaches $q_c$ at smaller
$\nabla \phi$ while in the second case $I_c$ may even slightly
increase at weak magnetic field (note that these fields weakly
affect superconductivity in S-layer due to much larger value of
$q_c$). Difference in $I_c^{\pm}$ provides the diode effect
(appearance of nonzero average voltage) in the regime with ac
current (with zero time-averaged current). In Fig. 5 we show
calculated dependence $I_c^{\pm}(H_{||})$ for bilayer with
following parameters: $d_N=d_S=2\xi_c$, $\rho_S/\rho_N=200$,
$T=0.2T_{c0}$. At large $H_{||}$ the superconductivity in N layer
is suppressed and $I_c^- \simeq I_c^+$.
\begin{figure}[hbtp]
\includegraphics[width=0.5\textwidth]{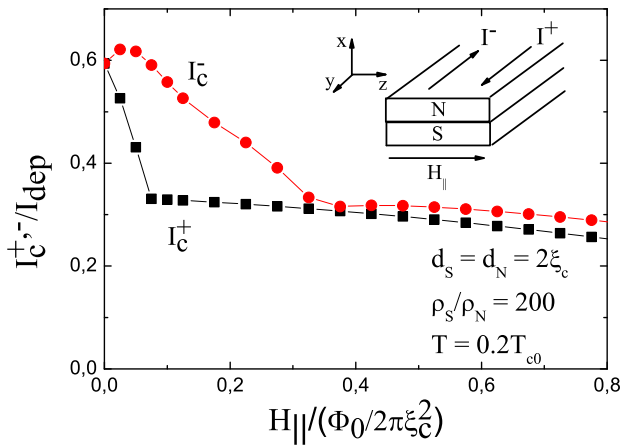}
\caption{Dependence of the critical current of SN bilayer on the
parallel magnetic field. Currents flowing in opposite directions
('positive' and 'negative' - see inset) have different critical
values when $H_{||} \neq 0$ which is a consequence of different
critical supervelocities in S and N layers.}
\end{figure}

Note that in symmetric SNS or NSN system the diode effect is
absent. It is rather weak in SN bilayer with $\rho_S \lesssim
\rho_N $ (comparable with one shown in Fig. 5 at large $H_{||}$)
because in this case the main part of superconducting current
flows via S-layer. Therefore the presence of noticeable diode
effect could be some kind of experimental verification of
large/dominant contribution of N-layer in $I_c$ at zero magnetic
field.

\section{Experimental results}

To verify theoretical predictions we perform experiments on
NbN/Al, NbN/Ag and MoN/Ag bilayers with $50
 \lesssim \rho_S/\rho_N \lesssim 400$ and single NbN, MoN films.
Using $T_{c0}$(NbN)=9 K and $T_{c0}$(MoN)=7.5 K we estimate
$\xi_c=6.5 nm$ for NbN (we take $D_S=0.5 cm^2/s$ from
\cite{Engel}) and $\xi_c=6.4 nm$ for MoN (we take $D_S=0.4 cm^2/s$
from \cite{Korneeva}). We have to mention, that critical
temperature of our NbN and MoN films gradually decreases with
decreasing their thickness when $d_S\lesssim 20 nm$ (results for
relatively thin MoN films are present in Ref. \cite{Korneeva}). We
relate this effect with presence of 'dead' nonsuperconducting
layer with thickness $2-3 nm$ at the interface with substrate.
Proximity effect with this layer could provide decreasing $T_c$
with decreasing $d_S$. Therefore we estimate the effective
'superconducting' thickness of our NbN and MoN in the range $11-13
nm$ which is close to $2 \xi_c$. In our experiment $\Lambda^{-1}$
is measured using two coils technique \cite{Claassen} (via
measurements of their mutual inductance $M$ with sample between
them) while $I_c$ is extracted from current voltage
characteristics of S and SN bridges. To study effect of the
thickness of N-layer on superconducting properties of bilayer we
change $d_N$ by consequent ion etching (for experimental details
see Appendix B).
\begin{figure}[hbtp]
\includegraphics[width=0.53\textwidth]{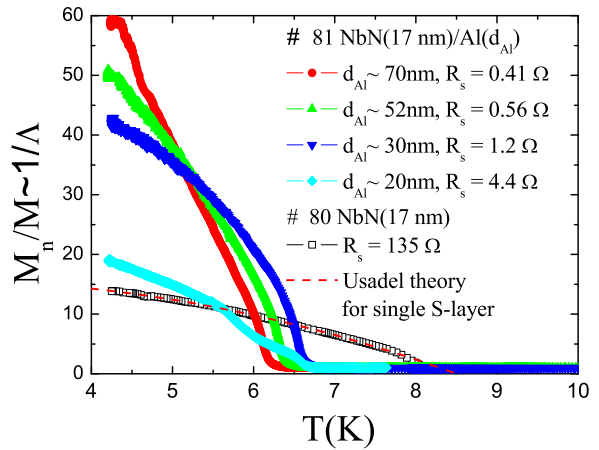}
\caption{Evolution of temperature dependence of $\Lambda^{-1}$ of
NbN/Al bilayer during etching and reference NbN film (measured
mutual inductance $M$ is normalized to its value in the normal
state $M_n$ just above $T_c$). Thickness of Al layer is estimated
from measured room temperature resistance per square $R_s$ (shown
in the inset). Red dashed curve corresponds to dirty limit
theoretical expectation for single superconducting layer:
$\Lambda^{-1} \sim \tanh(\Delta(T)/2k_BT)$ \cite{Tinkham_book}.}
\end{figure}

In Fig. 6 we show evolution of experimental $\Lambda^{-1}(T)$ for
NbN/Al bilayer during consequent etching of N-layer. One can see
that the presence of normal layer considerably enhances screening
abilities in comparison with the single superconducting film and
effect becomes stronger at low temperatures. With decreasing $d_N$
enhancement becomes smaller at low temperatures, which
qualitatively coincides with theoretical predictions (see Fig.
1(b)). Note that for thinnest Al layer shape of dependence
$\Lambda^{-1}(T)$ is probably affected by nonuniformity of Al
layer along the film (it appears during etching procedure and it
is seen from our measurements of resistance per square $R_s$ in
different places of the sample). Not etched bilayer NbN/Al with
similar $d_{\rm NbN}$ and $d_{\rm Al} = 10 nm$ shows much larger
diamagnetic response (see Fig. 12 in Appendix C) and no signs of
features in dependence $\Lambda^{-1}(T)$.
\begin{figure}[hbtp]
\includegraphics[width=0.46\textwidth]{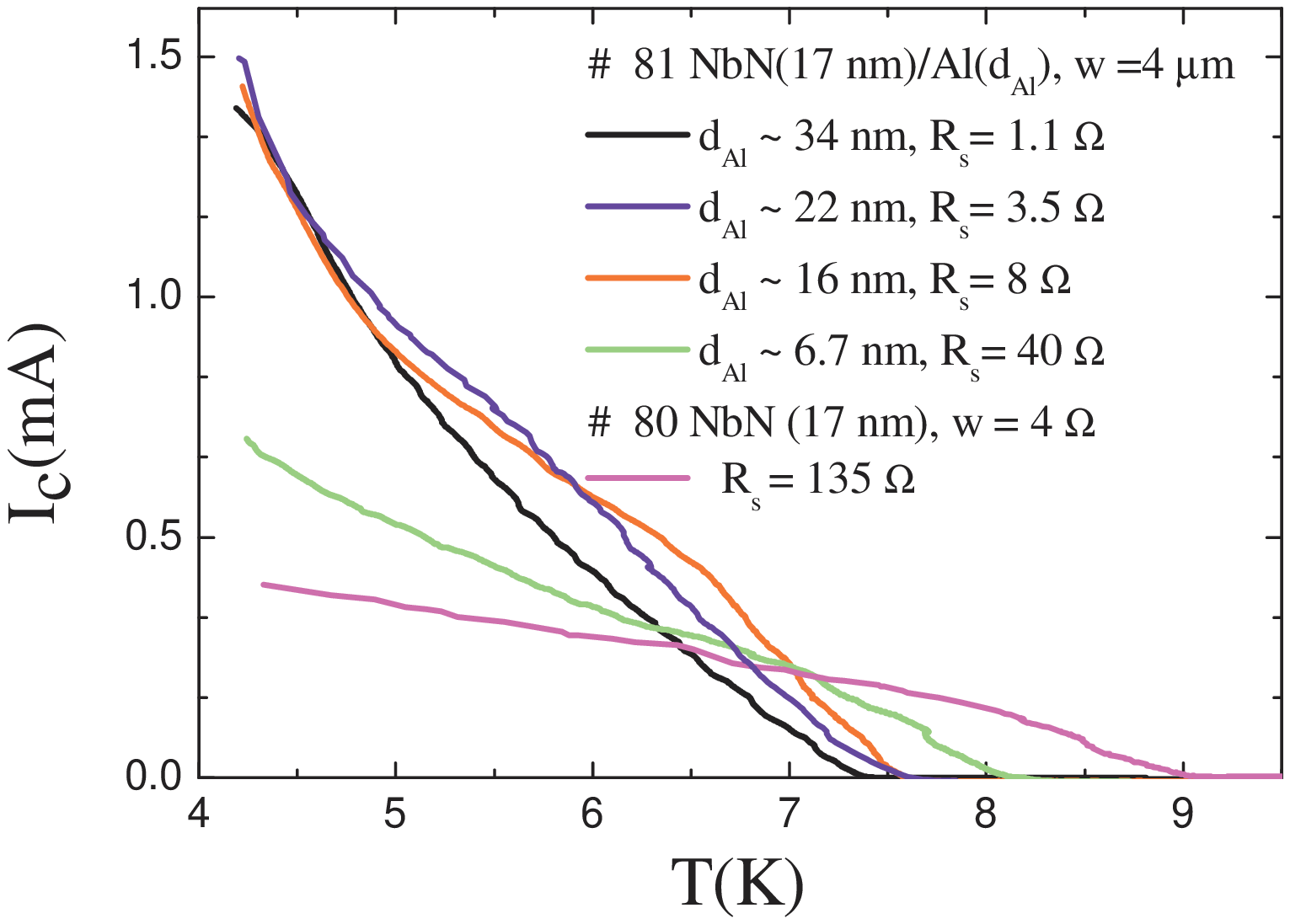}
\caption{Evolution of temperature dependence of critical current
of NbN/Al bridge during etching procedure. For comparison we also
show temperature dependence of critical current of NbN bridge.
Thicknesses of Al layers are estimated from room temperature $R_s$
(shown in the inset).}
\end{figure}

In Fig. 7 we show evolution of $I_c(T)$ of NbN/Al bridge with
width $w=4 \mu m$ during consequent etching (similar results are
found for NbN/Ag bridge - see Fig. 13 in Appendix C). Because
$\Lambda_0=\lambda(0)^2/d_S\simeq 21 \mu m$ (for estimation we use
$\lambda(0)=600 nm$ found from dirty limit expression
$\lambda(0)=(\hbar \rho/\pi \mu_0 1.76 k_BT_c)^{1/2}$ with
$\rho=300 \mu \Omega \cdot cm$ and $T_c=9 K$) and as it follows
from Fig. 12 for NbN/Al bridge
$\Lambda(4.2K)\simeq\Lambda_0/3\simeq 7 \mu m > w$ we conclude
that current flows uniformly across the bridge. One can see that
even few nanometer Al layer modifies $I_c(T)$ in comparison with
one for NbN bridge and makes it similar to the theoretical
expectations. We also find that $I_c$ of NbN/Al bridge with
relatively large $d_N$ exceeds critical current of NbN bridge in
about 4 times at $T=4.2 K$. So large critical current enhancement
we explain by low $I_c$ of our NbN bridge, in comparison with its
depairing current. Indeed, using theoretical expression following
from the Usadel theory (see for example Eq. (30) in \cite{Clem}),
measured $R_s(T=10 K)=150 \Omega$, $T_{c0}=9 K$ for sample $\#$80
and $D=0.5 cm^2/s$ we find $I_{dep}(4.2K)=3.8 mA$ which is 10
times larger than the measured critical current at T=4.2K. The
small critical current is probably connected with intrinsic
inhomogeneities of our NbN films which allow vortex penetration
and motion at supervelocity much lower than depairing
supervelocity (which corresponds to maxima in dependence $I_s(q)$
for ideal S-layer - see Fig. 2). In bilayer value of critical
supervelocity is mainly determined by N-layer (see maxima in
dependencies $I_s(q)$ in Fig. 2) and it is order of magnitude
smaller (for $\rho_S/\rho_N = 200$) than in ideal (defectless)
superconducting film. Because Al-layer is rather homogenous (this
fact follows from its relatively low resistivity and mean path
length $\ell_N \sim d_N$) we expect that at $q<q_c$ vortices
cannot penetrate our bilayer bridge. As a result the
superconducting current of bilayer bridge approaches its maximal
possible value (with main contribution from N-layer) which is
about half of $I_{dep}$ of NbN layer at this temperature
(according to our calculations - see Fig. 3 for bilayers with
close parameters).
\begin{figure}[hbtp]
\includegraphics[width=0.53\textwidth]{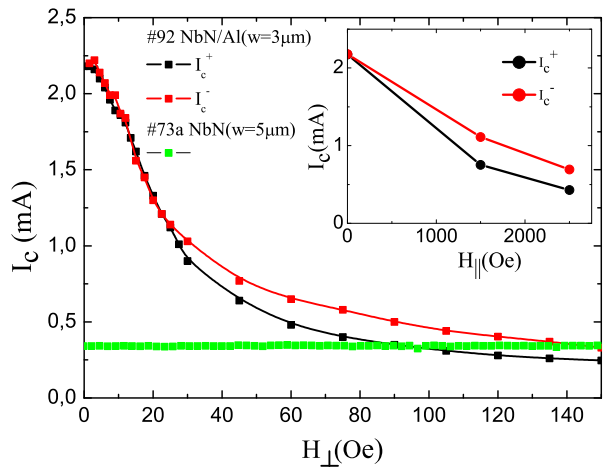}
\caption{Dependence of critical current (flowing in opposite
directions) of NbN/Al ($d_{\rm NbN}=19 nm$, $d_{\rm Al}=10 nm$)
and NbN ($d_{\rm NbN}=19 nm$) bridges on perpendicular magnetic
field $H_{\bot}$. For NbN bridge $I_c$ does not depend on
direction of the current and $H_{\bot}$. In the inset we show
dependence $I_c^{\pm}$ of the same NbN/Al bridge on the parallel
magnetic field.}
\end{figure}

Support for such an explanation comes from dependence of $I_c$ on
perpendicular magnetic field $H_{\bot}$ present in Fig. 8. In NbN
bridge $I_c$ does not depend on $H_{\bot}$ at weak magnetic fields
which confirms the idea that critical current is determined by
bulk pinning of the vortices with pinning current density $j_p \ll
j_{dep}$. Contrary, in NbN/Al bridge the shape of dependence
$I_c(H)$ says in favor of edge barrier controlled vortex
penetration and motion \cite{Plourde}. From dependence $I_c(H)$
one can estimate value of $q_c$ in N-layer and compare it with
corresponding value of single S-layer. From the edge barrier
theory it follows that at $H^* \simeq q_c \Phi_0/2\pi w$ critical
current drops in 2 times in comparison with $I_c(H=0)$. For 3 $\mu
m$ bridge $H^*\simeq 25 Oe$ (see Fig. 8) and $q_c\xi_c \simeq 0.15
$ which is much smaller than $q_c\xi_c \simeq 0.68$ for single S
layer at $T=0.5T_{c0}$.

Our measurements in parallel magnetic field revealed the diode
effect - value of the critical current depends on direction of its
flow (see inset in Fig. 8). In the experiment difference
$I_c^+-I_c^-$ is smaller than the theory predicts (compare Fig. 5
and inset in Fig. 8). We believe that it is connected with the
presence of perpendicular component of the magnetic field which
strongly suppresses proximity induced superconductivity in N-layer
and $I_c^{\pm}$. Note, that in the experiment $I_c^+ \neq I_c^-$
in the perpendicular magnetic field too (most probably it is
connected with different quality of edges of the bridge
\cite{Vodolazov}) but relative difference is smaller than in the
parallel magnetic field.

\section{Discussion}

We present results on superconducting properties of thin dirty SN
bilayer with thickness of S layer $d_S$ about several $\xi_c$,
thickness of N layer $d_N \ll \xi_N(T_{c0})$ and large ratio of
residual resistivities $\rho_S/\rho_N \gg 1$. We show that such a
bilayer has unique superconducting properties. First of all the
screening ability of the bilayer is determined mainly by the
proximity induced superconductivity in the normal layer, where
locally the London penetration depth $\lambda$ is smaller.
Secondly, at some conditions, the presence of normal layer may
considerably increase the critical current (the effect is not
connected with enhanced vortex pinning) because the largest part
of superconducting current flows via the normal layer where
superconducting current density $j_s\sim 1/\lambda^2$. Besides the
temperature dependence of critical current and effective magnetic
field penetration depth have unusual, non-BCS like temperature
dependence. We argue that these properties are consequences of
small thickness of N-layer and large ratio of resistivities
$\rho_S/\rho_N \gg 1$.
\begin{figure}[hbtp]
\includegraphics[width=0.52\textwidth]{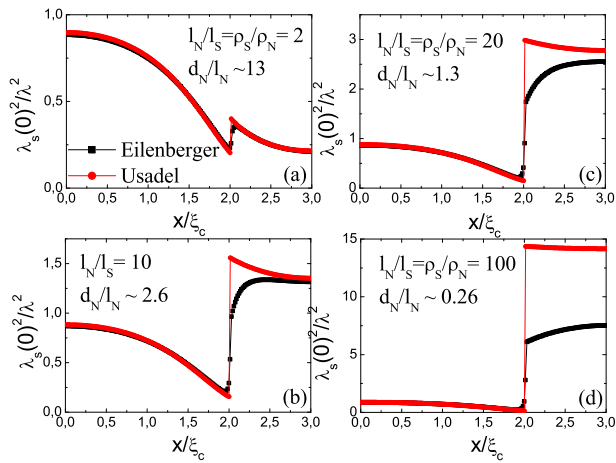}
\caption{Dependence of local $1/\lambda^2$ on the coordinate
across the bilayer ($d_S=2\xi_c$, $d_N=\xi_c$, $T=0.2 T_{c0}$)
calculated in Usadel and Eilenberger (see for example Eq. (1) in
\cite{Belzig_2}) models for different ratios of resistivities
(mean path lengths).}
\end{figure}

We believe that our results could be used as an alternative
explanation for the several times enhancement of $I_c$ found in
NbN/CuNi bilayers in comparison with single NbN film
\cite{Marrocco,Nasti}. The thickness of NbN layer was 8 nm, while
the thickness of CuNi layer varied from 3 up to 6 nm. It is know
that residual resistivity of CuNi strongly depends on Ni
concentration and varies in the range $\rho_{\rm CuNi} \simeq 2-49
\mu \Omega \cdot cm $ \cite{Ho} and, hence, potentially the ratio
$\rho_{\rm NbN}/\rho_{\rm CuNi}$ could reach 100 (with $\rho_{\rm
NbN}=200 \mu \Omega \cdot cm$). Unfortunately due to absence of
dependence $I_c(T)$ and actual values of resistivity of used CuNi
and NbN layers we cannot make a solid statement about the role of
proximity effect in that experiments.
\begin{figure}[hbtp]
\includegraphics[width=0.55\textwidth]{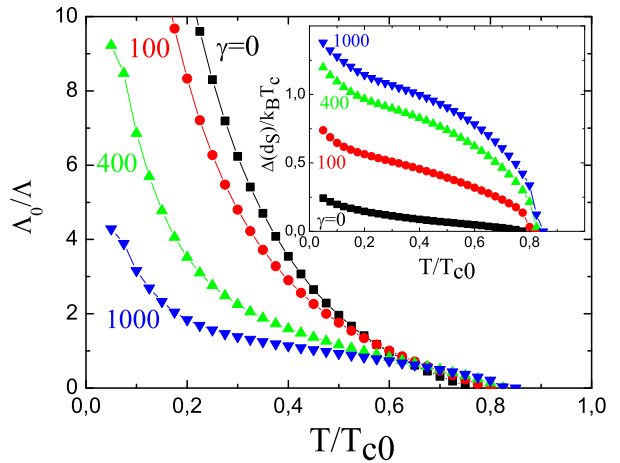}
\caption{Temperature dependence of $\Lambda^{-1}$ for bilayer with
$d_S=d_N=2\xi_c$, $\rho_S/\rho_N=200$ and finite barrier between S
and N layers (strength of the barrier is governed by parameter
$\gamma$  - see Eq. A6). In the inset we present temperature
dependence of $\Delta(d_S)$ at different $\gamma$. Result are
obtained using Usadel model.}
\end{figure}

Our experimental findings confirm main predictions of the theory
but there is quantitative discrepancy between theory and
experiment. The main difference is that in the experiment effect
of N-layer is stronger than Usadel model gives. First of all $T_c$
of bilayer changes with change of $d_N$ stronger and
$\Lambda^{-1}$ is larger in the experiment than in the theory
(compare Fig. 1(a) with Fig. 6 and Fig. 11 in Appendix C).
Secondly, there is a difference in the predicted shape of
dependence $\Lambda^{-1}(T)$ and experimental one for relatively
small $d_N$. And finally, critical current of the bilayer is
larger in the experiment. For example for sample $\#$92 ($d_{\rm
NbN}=19 nm$, $d_{\rm Al}=10 nm$, $w=3 \mu m$ - see Fig. 8)
experimental critical current of bilayer approaches $I_{dep}$ of
host superconductor already at $T=4.2 K=0.47 T_{c0}$ while theory
predicts only $I_c=0.29-0.31 I_{dep}$ depending on the chosen
ratio $\rho_S/\rho_N=50-200$ (in calculations we use $d_S=2\xi_c$
and $d_N=1.5 \xi_c$).

These quantitative discrepancies are not related to the usage of
Usadel theory in N-layer where mean path length $\ell_N \sim d_N$
or finite transparency of SN-interface (both of them lead to
decreasing of $\Lambda^{-1}$ - see Figs. 9 and 10). Even if one
takes into account dependence of $\rho_N$ on $d_N$ it provides
only a little better quantitative fit to the experiment. The
origin of the difference could be related with very short mean
path length $\ell_S$ in NbN and MoN materials which is in the
range of $ 1 \AA$ if one uses relation $D_S=v_F\ell_S/3$ with
typical value of Fermi velocity $v_F=2\cdot 10^8 cm^2/s$. It
questions about boundary conditions for quasiclassical Green
functions at SN interface between highly disordered superconductor
and relatively clean metal and how it affects both direct and
inverse proximity effects in N and S layers.

Due to strong dependence of $I_c$ on temperature such a SN bilayer
has unclear perspectives from point of view applications where one
needs large critical current. Besides effect of current
enhancement could be noticeable only in relatively thin structures
and when host superconductor has critical current much smaller
than depairing current. From another side steep temperature
dependence of $I_c$ and $\Lambda^{-1}$ could be utilized in
different kinds of superconducting detectors of electromagnetic
radiation \cite{Natarajan,Day}, particles \cite{Shishido} or dark
matter \cite{Hochberg} based on temperature dependent $I_c(T)$
and/or $\Lambda^{-1}(T)$. For example in superconducting single
photon detector (SSPD) absorbed photon or particle locally heats
by $\delta T$ the superconducting strip biased below its critical
current $I<I_c(T)$ and the superconductor transits to the
resistive state when $I>I_c(T+\delta T)$ \cite{Natarajan}. In the
kinetic inductance detector (KID) such a heating leads to change
of kinetic inductance $L \sim \Lambda^{-1}(T)$ and resonance
frequency of corresponding inductance-capacity circuit \cite{Day}.
It is clear that the steeper the temperature dependence of $I_c$,
$\Lambda^{-1}$ the larger will be their change at fixed $\delta T$
(it is determined by energy of absorbed particle or photon) and
sensitivity of detector should subsequently increase.

Another advantage of SN bilayer in comparison with highly
disordered superconductors is connected with high uniformity of
N-layer. Our results show that proximity induced superconductivity
in N-layer is weakly sensitive to local inhomogeneities of host
superconductor (it follows from our measurements of $I_c(H)$ for
SN bilayer - see Fig. 8) and critical current of bilayer
approaches to its maximal possible value. Note, that in
superconductors with large $\rho$ the value of critical current is
dictated by the weakest place in the sample and usually it is
smaller than the depairing current in two or more times
\cite{Lusche}.

\begin{acknowledgments}
The work is supported by the Russian Scientific Foundation, grant
No. 15-12-10020 (A.V.P., A.M.K.) and grant No. 17-72-30036
(D.Yu.V.). E.E.P., S.S.U., M. Yu. L. and A.I.E. acknowledge
support from Russian Foundation for Basic Research, grants No
15-42-02365 and No 15-42-02469.  The authors also thank N.V.
Rogozhkina and A.N. Kovtun for the help with the samples
fabrication. The facilities of the Common Research Center 'Physics
and Technology of Micro- and Nanostructures' of Institute for
Physics of Microstructures of RAS were used.
\end{acknowledgments}

\appendix

\section{Model}

To calculate superconducting properties of SN bilayer we use
Usadel equation for anomalous $F=\sin \Theta$ and normal $G=\cos
\Theta$ Green functions

\begin{equation}
\hbar D \frac{\partial^2\Theta}{\partial
x^2}-\left(2\hbar\omega_n+\frac{D}{\hbar}q^2\cos \Theta\right)\sin
\Theta+2\Delta \cos \Theta=0,
\end{equation}
where $D$ is a diffusion coefficient ($D=D_S$ in superconducting
layer and $D=D_N$ in the normal one), $\omega_n=\pi T(2n+1)$ is a
Matsubara frequency, $q=\nabla \varphi- (2\pi/\Phi_0)A(x)$
($\varphi$ is a phase of the order parameter, $A$ is a vector
potential) takes into account nonzero velocity of superconducting
condensate $v_s \sim q$ in direction parallel to layers ($y$
direction in our case), $\Delta(x)$ is a magnitude of
superconducting order parameter which has to be found in the
superconducting layer with help of self-consistency equation
\begin{equation}
\Delta \ln\left(\frac{T}{T_{c0}}\right)+2\pi k_B T\sum_{\omega_n
\geq 0} \left( \frac{\Delta}{\hbar\omega_n}-\sin \Theta_s\right)=0
\end{equation}
and we assume that in the normal layer $\Delta=0$ because of zero
BCS coupling constant. $T_{c0}$ in Eq.(A2) is the critical
temperature of superconductor with no N-layer.

We consider thin bilayer with thickness of superconducting layer
$d_S\ll \lambda$ ($\lambda$ is the London penetration depth) and
thickness of normal layer $d_N$ less than characteristic
penetration depth of magnetic field in N layer. Therefore we may
neglect corrections to $A(x)$ which comes from screening effect
and choose following form for $A(x)=H_{||}x$ ($H_{||}$ is a
parallel magnetic field).

The inverse effective magnetic field penetration depth by
definition is
\begin{equation}
\Lambda^{-1} =\frac{16 \pi^2}{\hbar c^2}
\int_0^{d_s+d_n}\frac{1}{\rho}\sum_{\omega_n \geq 0} \sin^2\Theta
dx
\end{equation}
with $\rho=\rho_S$ in S layer and $\rho=\rho_N$ in N layer. In the
absence of normal layer $\Lambda=\lambda^2/d_S$ - so called Pearl
penetration depth.

To find the sheet critical current (critical current per unit of
width of the bilayer) we use the following expression for sheet
superconducting current
\begin{equation}
J=\int_{0}^{d_s+d_n}\frac{2\pi k_BT}{e\hbar\rho} q\sum_{\omega_n
\geq 0} \sin^2\Theta dx
\end{equation}

The sheet critical current is defined as maximal sheet
superconducting current. For superconducting film without N-layer
$J_c=j_{dep}d_s$, where $j_{dep}$ is the depairing current.

At SN interface ($x=d_S$) we use Kupriyanov-Lukichev boundary
conditions \cite{KL}
\begin{equation}
\left.D_s\frac{d\Theta}{dx}\right|_{x=d_s-0}=\left.D_n\frac{d\Theta}{dx}\right|_{x=d_s+0}
\end{equation}

\begin{equation}
\gamma \xi_c
\left.\frac{d\Theta}{dx}\right|_{x=d_s+0}=\sin(\Theta(d_s+0)-\Theta(d_s-0))
\end{equation}
and boundary condition with vacuum at $x=0,d_s+d_n$:
$d\Theta/dx=0$. Eq. (A6) leads to jump of $\Theta$ on SN boundary
in presence of the barrier, which is controlled by parameter
$\gamma=R_{SN}A_{SN}/(\sigma_N\xi_c)$ ($R_{SN}$ is the resistance
of SN interface, $A_{SN}$ is its area and $\xi_c=\sqrt{\hbar
D_S/k_BT_{c0}}$). Usually we choose $\gamma=0$ which leads to
continuity of $\Theta$: $\Theta(d_S+0)=\Theta(d_S-0)$.

Equations (A1,A2) are solved numerically by using iteration
procedure. For initial distribution $\Delta(x)=const$ we solve Eq.
(A1) for Matsubara frequencies ranging from n=0 up to n=100. In
numerical procedure we use Newton method combined with tridiagonal
matrix algorithm. Found solution $\Theta(x)$ is inserted to Eq.
(A2) to find $\Delta(x)$ and than iterations repeat until the
relative change in $\Delta(x)$ between two iterations does not
exceed $10^{-8}$. Length is normalized in units of $\xi_c$, energy
is in units of $k_BT_{c0}$, current is in units of depairing
current of single S-layer with the thickness $d_S$, magnetic field
is in units of $H_0=\Phi_0/2\pi\xi_c^2$ and effective magnetic
field penetration depth is in units of
$\Lambda_0=\lambda^2(T=0)/d_S$. Usual step grid in S and N layers
is $\delta x=0.02 \xi_c$.

To decrease the number of free parameters we suggest that the
density of states in S and N layers are the same and ratio of
resistivities is equal to inverse ratio of diffusion constants or
mean path lengths $\rho_S/\rho_N=D_N/D_S=\ell_N/\ell_S$.

\section{Experimental details}

The bilayers NbN/Ag and NbN/Al were prepared on ${\rm Al_2O_3}$
$10\times 10 mm^2$ substrates in a magnetron vacuum machine
(Alcatel SCM-600) with a load-lock chamber. The thin films were
fabricated in a single deposition run at the substrates at ambient
temperatures. In total three targets were used: pure niobium
(99.9$\%$) as a superconducting material, Al ((99.99$\%$) and Ag
(99.99$\%$) as the normal metals. The design of the deposition
machine allows growth of the entire structure in one cycle without
disrupting the vacuum. This results in high-quality structures
with clean interfaces and strong proximity effect. NbN films were
deposited by reactive dc-magnetron sputtering in the Ar
(99.999$\%$) and N2 (99.999$\%$) gases mixed at the total pressure
of $7 \times 10^{-3}$ mbar with a residual pressure in the chamber
of about $1.5 \times 10^{-7}$ mbar. The deposition rate of the NbN
layers was 1.3 nm/s. The Ag and Al films were deposited by
rf-magnetron sputtering in the Ar at the pressure of $2 \times
10^{-2}$ mbar. The deposition rates of the Ag and Al layers were
from 1 to 3 nm/s.

MoN/Ag was fabricated by DC-magnetron sputtering on HV system AJA
ATC-2200 at room temperature. All samples were fabricated on a
silicon substrate (KDB-10) in one vacuum cycle and covered by Si
(10 nm) to protect of oxidation a top layer. MoN film was
deposited from metallic Mo target in $N_2$ atmosphere, see
\cite{Korneeva}. The base vacuum in the main chamber was about $2
\times 10^{-8}$ mbar. The working pressure was maintained at a
level of $2.6 \times 10^{-8}$ mbar. The rate of deposition of
layers was about 3-5 nm/min.

Measurements of $\Lambda^{-1}$ were performed by using standard
two-coil measurement technique of mutual inductance (see for
example Ref. \cite{Claassen}). The diameter of coils (2 mm) and
their height (4 mm) is smaller than the typical lateral size of
measured film ($10\times 10$ or $7\times 7$ mm), separation
between coils is $\sim$ 1 mm, while thickness of the bilayer
$d_s+d_n < 100 $ nm is smaller than London penetration depth. At
these parameters mutual inductance $M\sim \Lambda$ \cite{Claassen}
except temperatures close to $T_c$ where $\Lambda$ diverges.

The bridges made of bilayer films were fabricated by standard
lift-off lithography. Because of low substrate temperature we were
able to deposit the bilayers on the photoresist without its
apparent degradation. Widths of the bridges range from 3 up to 5
microns, while their length is fixed to 10 microns. Due to low
thickness of studied bilayer bridges ($d_{\rm NbN}=12-19 nm$,
$d_{\rm MoN}=19 nm$, $d_{\rm Al}=2-25 nm$, $d_{\rm Ag}=2-30 nm$)
current density distribution across the sample is expected to be
uniform because $w < \Lambda$. Transport measurements were
performed by four-probe method. During measurements we change the
current from large negative up to large positive values and in
this way we were able to find both the critical current $I_c$ (at
this current bridge switches from the superconducting state to the
resistive one) and retrapping current $I_r$ (when the bridge
switches back from the resistive to the superconducting state).

Resistivity of the samples was measured either using van Der Pauw
method and/or transport measurements. In this way we find
$\rho_{\rm NbN} = 260-380 {\rm \mu \Omega\cdot cm}$ depending on
the sample and $\rho_{\rm MoN}= 200 {\rm \mu \Omega\cdot cm}$ at
$T=10 K$. For both materials room temperature $\rho$ is smaller
than at $T=10 K$ which is common feature of highly disordered
metallic films. Our thickest Al film ($d_{\rm Al}=70 nm$) has
$\rho_{\rm Al}=2.9 {\rm \mu \Omega\cdot cm}$ while 90 nm thick Ag
film has $\rho_{\rm Ag}=1.5 {\rm \mu \Omega\cdot cm}$ (both at
room temperature) which are close to literature data \cite{Ag_Al}.
Residual resistance for these and thinner films is 1-4 times
smaller and gradually increases with decreasing $d_N$ \cite{Ag_Al}
which gives us $\rho_S/\rho_N \lesssim 400$ depending on the
thickness of the normal layer and the pair of S,N materials.
\begin{figure}[hbtp]
\includegraphics[width=0.5\textwidth]{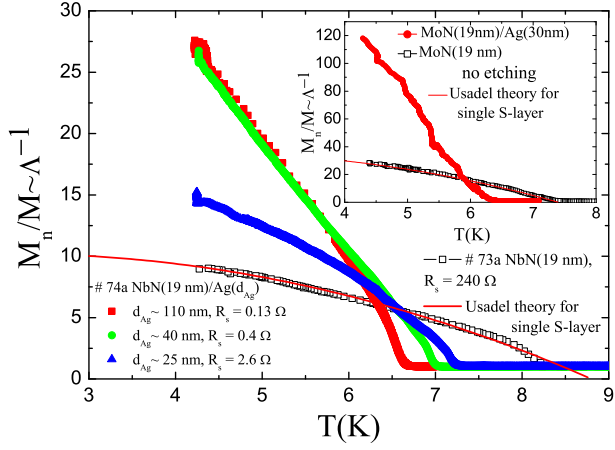}
\caption{Evolution of temperature dependence of $\Lambda^{-1}$ of
NbN/Ag bilayer after consequent etching and reference NbN film
(mutual inductance is normalized to its value in the normal state
just above $T_c$). In the inset we present $\Lambda^{-1}(T)$ for
MoN/Ag bilayer ($R_s=0.73 \Omega$ at room temperature) with
$d_{\rm MoN}=19 nm$, $d_{\rm Ag}=30 nm$ and single MoN layer
($R_s=120 \Omega$) with $d_{\rm MoN}=19 nm$.}
\end{figure}

To study effect of the thickness of normal layer on
superconducting properties of bilayer we make chips with side
$10mm \times 10 mm$. The central part of chips with size $6 mm
\times 6 mm$ is used for measurements of mutual inductance, while
on the edge of the chip we fabricate bridges. The thickness of Al
and Ag layers is changed by gradual ion etching. The ion etching
of Al and Ag layers was performed in a Plasmalab 80 plus (Oxford
Instruments) equipped with capacitive (HF) and inductive (ICP)
plasma sources. The etching process was performed in Ar at the
pressure of 5 mbar, HF power 100 W and ICP power 400 W. The Ag
etching rate was 30nm/min and Al etching rate 1 nm/min.
\begin{figure}[hbtp]
\includegraphics[width=0.5\textwidth]{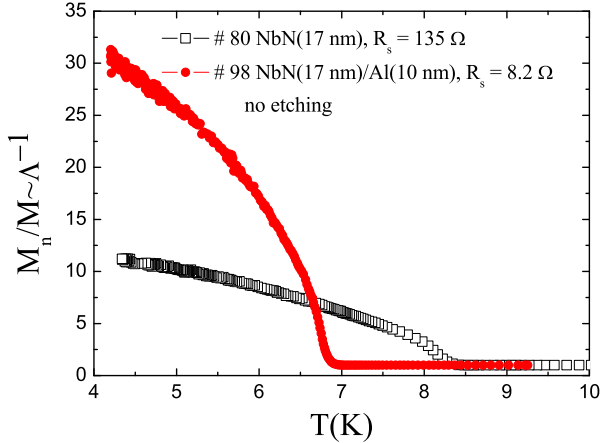}
\caption{Temperature dependence of $\Lambda^{-1}$ of NbN/Al
bilayer (sample $\#$98, $d_{\rm NbN}=17$ nm, $d_{\rm Al} \simeq
10$ nm, room temperature $R_s=8.2 \Omega$) and NbN film (sample
$\#$80, $d_{\rm NbN}=17$ nm, room temperature $R_s=135 \Omega$).}
\end{figure}

We also did special experiment and find dependence of resistance
per square $R_s$ of normal layer on its thickness at room
temperature during etching procedure combined with simultaneous
measurement of the resistance of the etched sample. We use these
results to estimate thicknesses of the Al and Ag layers in bilayer
by measuring their room temperature $R_s$.

\section{Experimental results for NbN/Ag and MoN/Ag bilayers}

In Fig. 11 we present experimental $\Lambda^{-1}(T)$ for NbN/Ag
bilayer after two consequent etching of N-layer. In the inset to
Fig. 11 we show results for MoN/Ag bilayer and in Fig. 12 for
NbN/Al bilayer without etching. Qualitatively these results
coincide with ones for NbN/Al bilayers and theoretical
calculations in the Usadel model. Quantitatively in the experiment
$\Lambda^{-1}$ is larger than the theory predicts.

\begin{figure}[hbtp]
\includegraphics[width=0.47\textwidth]{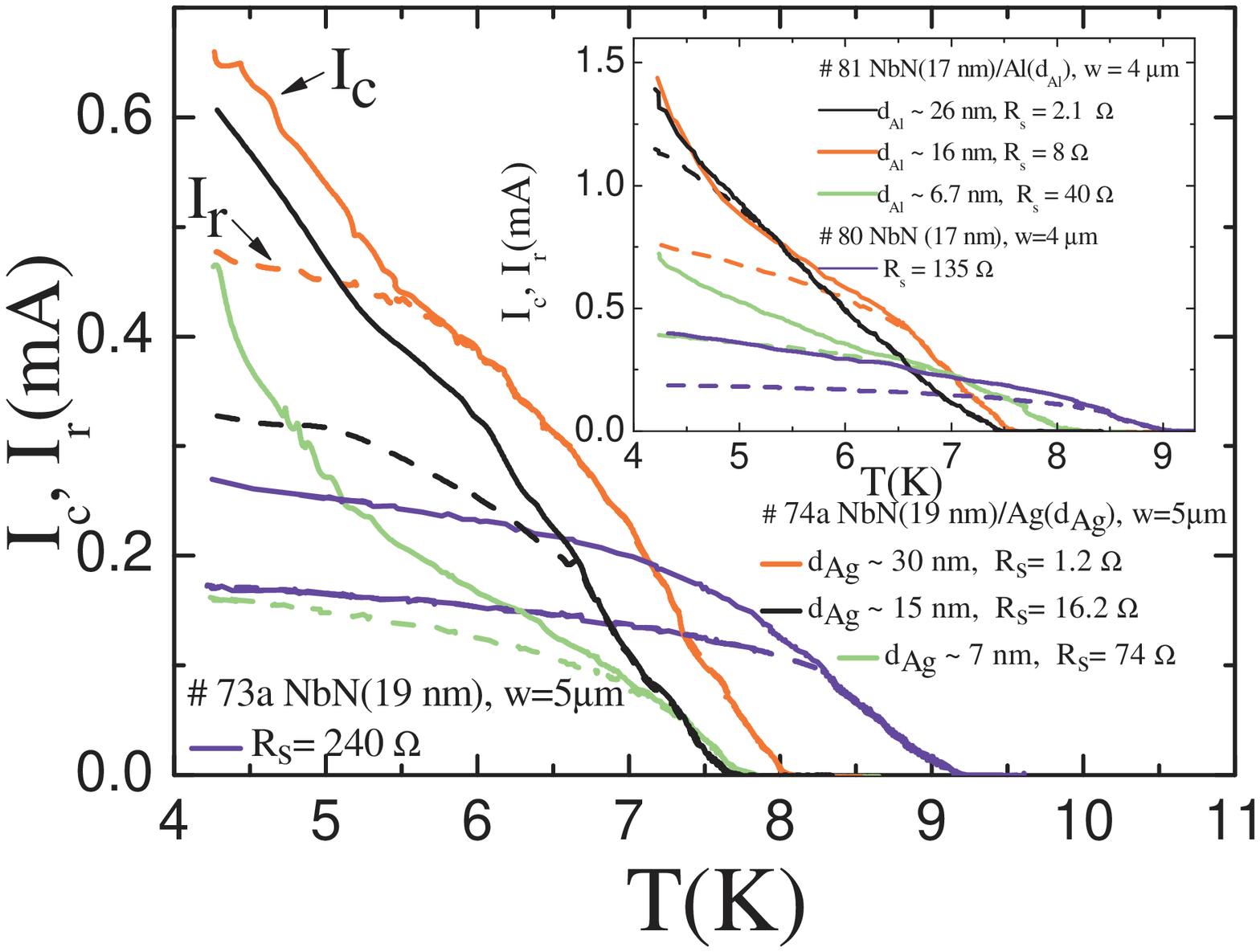}
\caption{Temperature dependence of critical (solid curves) and
retrapping (dashed curves) currents of NbN/Ag and NbN/Al bridges
(see inset) with different thicknesses of Al and Ag layers. In the
same figures we show $I_c(T)$ and $I_r(T)$ of reference NbN
bridges.}
\end{figure}

In Fig. 13 we show temperature dependence of critical $I_c$ and
retrapping $I_r$ currents of NbN/Ag bridge with width $w=5 \mu m$
during consequent etching (in the inset results for NbN/Al bridge
with $w=4 \mu m$ are present). The presence on the normal layer
not only considerably increases the critical current and changes
its temperature dependence in comparison with superconducting
bridge but it also wipes out the hysteresis of current-voltage
characteristics (making $I_r=I_c$) for bilayers with relatively
large $d_N$. The last effect was observed earlier for shunted MoGe
superconducting bridge with low shunt resistance \cite{Brenner}.
The dependence $I_c(T)$ looks more 'noisy' for NbN/Ag bridges than
for NbN/Al ones. The origin of the 'noise' comes from stochastic
nature of switching to the resistive state \cite{Brenner} and it
is clear seen in our experiment via repeated measurements of
current-voltage characteristics at fixed temperature.

\end{document}